\documentclass[a4paper,aps,prl,twocolumn,noshowpacs,preprintnumbers,nofootinbib]{revtex4-1}

\usepackage[utf8x]{inputenc}
\usepackage{epsfig,amssymb,amsmath,psfrag}

\usepackage{etex}
\usepackage{pstricks-add}

\usepackage{bm}
\usepackage{epic}
\usepackage{graphicx}
\usepackage{color}
\usepackage{hyperref}
\usepackage{young}
\usepackage{multirow}
\usepackage{fancybox}

\usepackage[paperwidth=210mm,paperheight=297mm,centering,hmargin=1.75cm,vmargin=2.5cm]{geometry}

\def \be  {\begin{equation}}
\def \ee  {\end{equation}}
\def \ba  {\begin{eqnarray}}
\def \ea  {\end{eqnarray}}
\def \baa {\begin{eqnarray*}}
\def \eaa {\end{eqnarray*}}
\def \bb  {\begin {thebibliography} }
\def \eb  {\end{thebibliography}}
\def \lab #1 {\label{#1}}

\newcommand{\beq}{\begin{equation}}
\newcommand{\eeq}{\end{equation}}
\newcommand{\beqa}{\begin{eqnarray}}
\newcommand{\eeqa}{\end{eqnarray}}

\begin{document}

\preprint{QMUL-PH-16-10}

\title{An exact limit of ABJM}

\author{Marco S. Bianchi${}^{1}$}
\author{Matias Leoni${}^{2}$}

\affiliation{
${}^1${
Centre for Research in String Theory,
School of Physics and Astronomy,
Queen Mary University of London,
Mile End Road, London E1 4NS, UK}\\
${}^2${Physics Department, FCEyN-UBA \& IFIBA-CONICET\
Ciudad Universitaria, Pabell\'on I, 1428, Buenos Aires, Argentina}\\
{\tt m.s.bianchi@qmul.ac.uk, leoni@df.uba.ar}}
\begin{abstract}
We study planar ABJM in a limit where one coupling is negligible compared to the other. We provide a recipe for exactly solving the expectation value of bosonic BPS Wilson loops on arbitrary smooth contours, or the leading divergence for cusped ones, using results from localization. As an application, we compute the exact (generalized) cusp anomalous dimension and Bremsstrahlung function and use it to determine the interpolating $h$-function. We finally prove a conjecture on the exact form of the dilatation operator in a closed sector, hinting at the integrability of this limit.
\end{abstract}

\maketitle

\section{Introduction}
\label{sec:introduction}

ABJM is a three-dimensional ${\cal N}=6$ superconformal theory with a Chern-Simons action for the gauge groups $U(N)_k\times U(N)_{-k}$ and bifundamental matter \cite{ABJM}.
As for ${\cal N}=4$ SYM in four dimensions, this theory possesses a gravity dual at strong coupling via the AdS/CFT correspondence \cite{Maldacena:1997re} and several exact results are available for it. On the one hand the model is believed  \cite{MZ,GGY,GV,BR,BGR} to be integrable in the planar limit \cite{Beisert:2010jr}, despite of the appearance of an interpolating function of the coupling, whose exact expression has been conjectured in \cite{Gromov:2014eha}. On the other hand the theory can be localized \cite{Pestun:2007rz}, which allows for an exact evaluation of the expectation values of supersymmetric Wilson loop operators \cite{Kapustin:2009kz,Marino:2009jd,Drukker:2010nc}.

In this paper we consider ABJM theory with different ranks for the gauge groups $U(N_1)_k\times U(N_2)_{-k}$, also referred to as the ABJ model \cite{ABJ}.
We focus on its planar limit, where we define the effective 't Hooft couplings 
$\lambda_i \equiv \frac{N_i}{k}$.
We consider the limit where one coupling is negligible compared to the other, namely $\lambda_1\ll \lambda_2$ \cite{Minahan:2010nn}. 
We refer to such a limit as {\it extremal} ABJ.
It was argued \cite{ABJ} that unitarity imposes the bound $|N_1-N_2|<k$. This can be satisfied in the extremal case by taking $k$ sufficiently large, namely in the perturbative regime.

We claim that in the extremal limit the expectation values of certain Wilson loops can be computed exactly.
We arrive at this conclusion by the following chain of reasoning, on which we elaborate in the following sections. First, a Feynman diagram analysis of such a computation reveals that only a restricted class of diagrams contributes in this limit, namely the (quantum corrected) two-point functions of the connections.
Next, we analyze the matrix model average computing the 1/6-BPS circular Wilson loop via localization. We provide an ansatz solving it perturbatively in the extremal limit and find the exact result for the Wilson loop. Comparing its expectation value to the perturbative computation, we are able to extract exact expressions for the two-point functions of the connections. The knowledge of these building blocks then allows to compute the expectation value of all Wilson loops on arbitrary contours.

As an application we compute the exact (generalized) cusp anomalous dimension and Bremsstrahlung function \cite{Correa:2012at} in the extremal limit.
These are central objects of the theory, which provide a connection to integrability \cite{Beisert:2006ez,Correa:2012hh,Drukker:2012de}.
There are doubts on whether the ABJ model is integrable or not. In this paper we provide strong indications that at least in the extremal limit the theory indeed benefits from integrability.
Indeed we argue via superfield diagrammatics that in the extremal limit, in a closed subsector, the dilatation operator has the form of two decoupled Heisenberg spin chains to all loops, as conjectured in \cite{Minahan:2010nn}.
Moreover, using the cusp anomalous dimension result we also fix the interpolating $h$-function appearing in front of the dilatation operator, proving the full exact form conjectured in \cite{Minahan:2010nn}.

\section{Wilson loops in the extremal limit}
\label{sec:WL}

We consider bosonic Wilson loop operators for the $U(N_1)$ gauge group of the form
\begin{equation}
\label{eq:WL}
W[C]= \frac{1}{N_1}\, {\rm Tr}\, P\, \exp{
\left( - i \int_C d\tau\, {\cal L}(\tau) \right) } \,.
\end{equation}
The connection for the ordinary operator is ${\cal L}= A_\mu \dot{x}^\mu$, $A_{\mu}$ being the $U(N_1)$ gauge field. For the 1/6-BPS it reads
${\cal L}_{1/6} = A_\mu \dot{x}^\mu - \frac{2\pi i}{k}  |\dot{x}| O$, where $O={\mathcal M}_J^{\; I} C_I \bar{C}^J$, the fields $C$, $\bar C$ are the bifundamental scalars and ${\mathcal M}={\rm diag}(-1,-1,1,1)$ \cite{DPY,Chen:2008bp,Rey:2008bh}. The contour $C$ is an arbitrary curve in $\mathbb{R}^3$ parametrized by $\tau$.
We recall that the BPS Wilson loop is finite for smooth contours, thanks to supersymmetry.
One can also define Wilson loops for the $U(N_2)$ connection. In the extremal limit $\lambda_1\ll\lambda_2$ their expectation value reduces at leading order to the pure Chern-Simons result \cite{Drukker:2010nc} and we will not be interested in them in this paper.

When evaluating expectation values in a perturbative expansion for $\lambda_1\ll\lambda_2\ll1$, one computes correlation functions of the objects appearing in the connections, namely the gauge vectors $A_{\mu}$ and the scalar bilinears $O$.
We enforce the extremal limit by keeping only the first nontrivial order in $\lambda_1$ in the expectation values of the Wilson loops, which means only one power of this coupling.
Then, in the planar limit, it is easy to see that since both the gauge vector and the scalar bilinear transform in the adjoint of $U(N_1)$, the perturbative expansion of the expectation values is truncated to their (color stripped) two-point functions only
\begin{equation}\label{eq:limitW}
\langle W \rangle = 1 - N_1 \int_{\tau_1>\tau_2}\, \langle {\cal L}(\tau_1) {\cal L}(\tau_2) \rangle + {\cal O}(\lambda_1^2) \,.
\end{equation}
We note that the same logic applies also to arbitrary correlation functions of Wilson loop operators.
In the extremal limit the two-point functions have a restricted class of planar quantum corrections, which do not generate further powers of $\lambda_1$.
Yet they are still nontrivial functions of the $\lambda_2$ coupling.
For the gauge propagator, gauge invariance fixes the form of the quantum corrections to possess the form, in Feynman gauge, \cite{Chen:1992ee}
\begin{align}
& \langle A_\mu(x) A_\nu(y) \rangle = 
f_{CS}(\lambda_i)\, \frac{i}{k}\, \varepsilon_{\mu\nu\rho} \frac{(x-y)^\rho}{|x-y|^3 } + \nonumber\\&~~ + f_{YM}(\lambda_i)\, \frac{1}{k}\, 
\left[ \frac{\delta_{\mu\nu}}{ |x- y|^2} +\dots \right] \,,
\end{align}
where the ellipsis stands for a total derivative term which vanishes in all the computation of this paper.
The functions $f_{CS}$ and $f_{YM}$ occur at even and odd loop order, respectively and their indices stand for the Chern-Simons and Yang-Mills tensor structure of their propagators.
These are generated by the geometric sum of all the 1PI contributions. In the extremal limit, the latter are in turn a bubble of scalars or fermions with all possible Chern-Simons interactions of the $U(N_2)$ gauge group inside.

The quantum corrections to the scalar bilinear two-point function are all 1PI thanks to the tracelessness of ${\cal M}$. Their backbone is basically given by the scalar bubble or a sequence of an odd number of alternating scalar and fermion bubbles joint by quartic Yukawa interactions. On top of this all possible $U(N_2)$ Chern-Simons quantum corrections inside the bubbles contribute in the extremal limit. Finally, mixed two-point functions of a gauge vector with a scalar bilinear are forbidden since ${\cal M}$ is traceless.

\section{Localization result in the extremal limit}
\label{sec:localization}

ABJ theory can be localized on $S^3$ and its partition function is given by the matrix model
\begin{align}
& Z(N_1, N_2, k)=\int\prod_{i=1}^{N_1} d\mu_i \prod_{j=1}^{N_2} d \nu_j \prod_{i<j}  \sinh^2 \frac{\mu_i -\mu_j}{2} \nonumber\\& \times \sinh^2 \frac{\nu_i -\nu_j}{2} 
\prod_{i,j} \cosh^2 \frac{\mu_i -\nu_j}{2}\,  e^{-\frac{k}{4\pi\, i}\left(  \sum_i \mu_i^2 +\sum_j \nu_j^2\right)}\nonumber \,.
\end{align}
The expectation value of the 1/6-BPS Wilson loop can be computed exactly as an average in this matrix model, whose solution is nevertheless nontrivial.
In particular it was shown that it can be computed via the integral ($t_1=2\pi i \lambda_1$, $t_2=-2\pi i \lambda_2$) \cite{Marino:2009jd}
\begin{equation}
\langle W_{1/6} \rangle=\frac{1}{\pi t_1} I_1, \quad  I_1=\int\limits_{1/a}^a  \tan^{-1} \sqrt{\frac{\alpha X -1-X^2  }{\beta X+1+X^2 }} d X\nonumber \,,
\end{equation}
where $\alpha\equiv a+\frac{1}{a}$, $\beta\equiv b+\frac{1}{b}$ and $a$, $b$ are the endpoints of the cuts around which the eigenvalues condense in the large $N$ limit. Their expressions in terms of the couplings in the weak regime can be obtained inverting perturbatively a map specified in \cite{Marino:2009jd}.
In the extremal limit the expectation value is at most linear in $t_1$ and hence it is easier to compute the derivative
\begin{equation}
\frac{\partial I_1}{\partial t_1}  = \frac{\partial I_1}{\partial \zeta} \frac{\partial \zeta}{\partial t_1} +  2 \frac{\partial I_1}{\partial \xi} \xi e^{t_1+t_2} \,.
\end{equation}
In turn one finds
\begin{align}
\frac{\partial I_1}{\partial \zeta}&=
- \frac{1}{\sqrt{ab}\,(1+ab)}\left(a\,K(k)-(a+b)\,\Pi(n|k)\right),\\
\frac{\partial I_1}{\partial \xi}&= \frac{\sqrt{ab}}{a+b}\,E(k) \,,
\label{I_1_zeta}
\end{align}
where $\Pi(n|k)$ is the complete elliptic integral of the third kind, $K(k), E(k)$ are elliptic integrals of the first and second kind, respectively, and the modulus and characteristic are given by 
\begin{equation}
\label{modulus}
k^2=\frac{(a^2-1)(b^2-1) }{(1+a b)^2} \,,\qquad 
n=\frac{b }{a} \frac{a^2-1}{1+a b}\nonumber \,.
\end{equation}
Finally the parameters are defined as
\begin{equation}
\zeta=\frac12\left(\alpha-\beta\right), \qquad \xi=\frac12\left(\alpha+\beta\right)\nonumber \,.
\end{equation}
We have verified that the following ansatz up to $O(t_1^3)$ terms
\begin{align}
\alpha &= 2+4t_1 e^{\frac{t_2}{2}}t_1 + \frac12(4e^{\frac{t_2}{2}}+e^{t_2}-1)t_1^2 \nonumber\\
\beta &= -2 +4 e^{t_2} + 4(e^{t_2}-e^{\frac{t_2}{2}})t_1 + \frac12(1-4e^{\frac{t_2}{2}}+3e^{t_2})t_1^2\nonumber
\end{align}
is a solution of the inversion problem for $a$ and $b$ up to the specified order in the extremal limit. The ${\cal O}(t_1^2)$ terms are needed to obtain the expectation value to the desired order.
Following the steps of \cite{Marino:2009jd} and integrating in $t_1$ we finally find
\begin{equation}
\langle W_{1/6} \rangle = 1 + \pi\, i\, \lambda_1\, e^{-\pi\, i\, \lambda_2} + {\cal O}(\lambda_1^2) \,.
\end{equation}
One can also compute the expectation value of the 1/6-BPS wound $m$ times around the great circle \cite{Marino:2009jd,Klemm:2012ii}.
From the explicit result of \cite{Bianchi:2016gpg}, we find indication that in the extremal limit
\begin{equation}
\langle W_{1/6}^m \rangle = 1 + \pi\, i\, \lambda_1\, e^{-\pi\, i\, \lambda_2}\, m^2 + {\cal O}(\lambda_1^2) \,.
\end{equation}
The fact that the winding number appears only with power $m^2$ is in perfect agreement with the perturbative insight that only two-point functions contribute to the expectation value in the extremal limit.

\section{Exact structures in extremal ABJ}
\label{sec:structures}

We interpret the localization result for extremal ABJ in light of the perturbative structure \eqref{eq:limitW} applied to the circular Wilson loop.
At odd loops only the quantum corrected gauge propagator contributes, since
corrections to the scalar bilinear two-point function vanish identically, thanks to the following argument. 
The Feynman rules imply that at odd loops an odd number of $\varepsilon$ tensors is generated. These can always be reduced to a single antisymmetric tensor, whose indices are contracted with derivatives acting on an integral which only depends on the vector $(x_1-x_2)^{\mu}$. The contraction then vanishes by antisymmetry.
At odd loops the gauge contributions are proportional to the Gauss integral
\begin{equation}
\frac{1}{4\pi} \oint_{C} dx_1^\mu dx_2^\nu \, \varepsilon_{\mu\nu\rho} \frac{(x_1-x_2)^\rho}{|x_1-x_2|^3} = n \,, \qquad n\in \mathbb{Z} \,.
\end{equation}
The latter vanishes for a planar contour as the circle, but is instead proportional to an integer number $n$ if a nontrivial framing of the path is specified \cite{Witten}.
Comparing this with the localization result, we ascertain that it has been derived at framing $n=-1$ and we find
\begin{equation}
f_{CS}(\lambda_2) = \frac{1}{2}\, \cos \pi\lambda_2 \,. 
\label{eq:exactprop1} 
\end{equation}
This induces a non-trivial dependence of the effect of framing on the coupling in ABJ, at a difference with respect to the pure Chern-Simons case, and  in agreement with the perturbative findings of \cite{Bianchi:2016yzj}.

At even loops two structures contribute, namely the quantum corrected propagator and the two-point function of the scalar bilinear $O$.
These are two separated objects, contributing with the integrands
\begin{align}\label{eq:struct1}
\raisebox{-0.55cm}{\includegraphics[width=1.25cm]{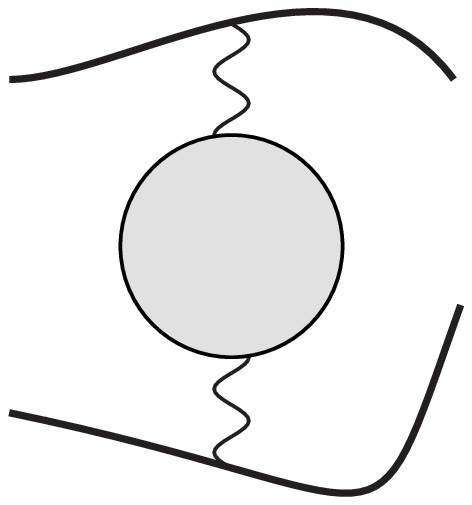}} &= \lambda_1\, f_{YM}(\lambda_2)\, \frac{-\dot x_1\cdot \dot x_2}{|x_1-x_2|^2} + {\cal O}(\lambda_1)\\
\raisebox{-0.55cm}{\includegraphics[width=1.25cm]{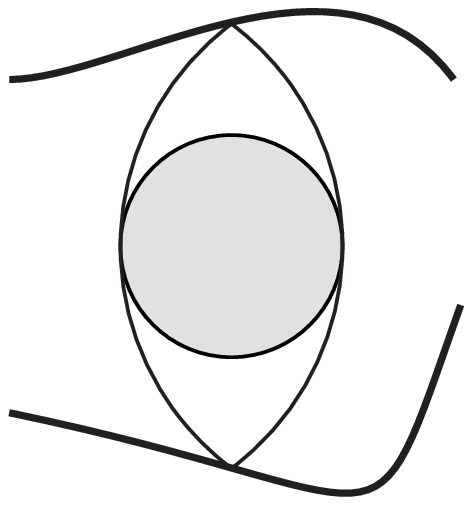}} &= \lambda_1\, f_{O}(\lambda_2)\, \frac{|\dot x_1||\dot x_2|}{|x_1-x_2|^2} + {\cal O}(\lambda_1) \,,
\label{eq:struct2}
\end{align}
where the coefficients of the structures are a priori two unrelated functions of $\lambda_2$.
We point out that the quantum corrections to the scalar two-point function are finite. Indeed this is practically the same computation as for the two-point function of the operator ${\rm Tr} O$, which is protected (being the matrix ${\cal M}$ traceless) and hence should not have an anomalous dimension.
Then we analyze separately the contour integrals of \eqref{eq:struct1} and \eqref{eq:struct2} on the straight line. In this case the spacetime structures coincide and the resulting contour integral is ultraviolet divergent.
Since $f_O$ is finite and the expectation value of the 1/6-BPS Wilson loop on the line is trivial, we conclude that the coefficients must be equal and combine into the integrand
\begin{equation}\label{eq:integrand}
\raisebox{-0.55cm}{\includegraphics[width=1.25cm]{gaugecorrection}} + \raisebox{-0.55cm}{\includegraphics[width=1.25cm]{scalarcorrection}} = \lambda_1 f_{YM}(\lambda_2) \frac{|\dot x_1||\dot x_2|-\dot x_1\cdot \dot x_2}{|x_1-x_2|^2} \,, 
\end{equation}
where the divergence at coincident points is mitigated by the vanishing of the numerator, leading to a finite integral for an arbitrary smooth contour.
Comparing the localization result with the the integral of \eqref{eq:integrand} on the circle we find
\begin{equation}
f_O(\lambda_2) = f_{YM}(\lambda_2) = \frac{1}{\pi}\, \sin \pi\lambda_2 \,. 
\label{eq:exactprop2} 
\end{equation}
We have also successfully tested this statement by computing the two-loop corrections to the scalar two-point function. Indeed, using partial results from \cite{MOS2} one can check that
\begin{equation}
\frac{1}{N_1}\, \left(\frac{2\pi}{k}\right)^2\, {\rm Tr} \langle O(x_1) O(x_2) \rangle^{(2)} = -\frac{\pi^2}{6} \frac{\lambda_1\lambda_2^3}{|x_1-x_2|^2}\nonumber \,,
\end{equation}
in agreement with \eqref{eq:exactprop2}.
We stress that the results \eqref{eq:exactprop1} and \eqref{eq:exactprop2} are gauge dependent and only valid in strictly three dimensions. Moreover an implicit choice of regularization might be implied in their derivation. Hence they cannot be employed in other computations using different gauges or regularization schemes.
Yet they are perfectly suitable for evaluating other Wilson loops expectation values in the extremal limit, provided the contour integration is finite, or one only focuses on the leading divergence.

\section{Exact Wilson loops}
\label{sec:exactWL}

Using the results from the previous sections and in particular \eqref{eq:exactprop1} and \eqref{eq:exactprop2} we are able to formulate the main statements of the paper.
We claim that in the extremal limit the exact expectation value of the ordinary Wilson loop on a generic contour is given by
\begin{align}
\langle W \rangle &= 1 - \lambda_1\, \bigg[ 2\pi\, i\, n\, f_{CS}(\lambda_2) + \nonumber\\& + \int_{\tau_1>\tau_2}\, \frac{\dot x_1\cdot \dot x_2}{|x_1-x_2|^2}\, f_{YM}(\lambda_2) \bigg] + {\cal O}(\lambda_1^2) \,,
\end{align}
whereas the expectation value of the 1/6-BPS Wilson loop evaluates
\begin{align}\label{eq:exactWL}
\langle W_{1/6} \rangle &= 1 - \lambda_1\, \bigg[ 2\pi\, i\, n\, f_{CS}(\lambda_2) + \\& +  \int_{\tau_1>\tau_2}\, \frac{\dot x_1\cdot \dot x_2-|\dot x_1||\dot x_2|}{|x_1-x_2|^2}\, f_{YM}(\lambda_2) \bigg] + {\cal O}(\lambda_1^2) \,, \nonumber
\end{align}
allowing for a framing of the path.
Similar exact formulae apply for all correlation functions of Wilson loops, at leading nontrivial order in the extremal limit.
We recall that the results above require the finiteness of the Wilson loop expectation value to hold. This is true for the 1/6-BPS Wilson loop as long as it is evaluated on a smooth (not light-like) path. For the ordinary Wilson loop this depends on the particular path. It is for instance finite on a circle (if evaluated in dimensional regularization \cite{Bianchi:2013zda,Bianchi:2013rma,Griguolo:2013sma}) but divergent on a straight line.
Nevertheless the technique above can also be used for divergent objects, if one focusses on the coefficient of the leading divergence.
As a particularly interesting example of such a situation we compute the cusp anomalous dimension of extremal ABJ. Using the formulae above and the computation of \cite{HPW,CH,BLMPS1,BLMPS2} we find the exact result 
\begin{equation}
\label{eq:cusp}
\Gamma_{cusp} = 4\, \lambda_1\, \frac{\sin\pi\lambda_2}{\pi} + {\cal O}(\lambda_1^2) \,.
\end{equation}
The prescription \eqref{eq:exactWL} can be adapted to cases with different coupling matrices ${\cal M}$, for instance in the configuration of the generalized cusp \cite{Griguolo:2012iq}. It suffices inserting a factor $\frac{{\rm Tr}({\cal M}_1{\cal M}_2)}{4}$ in front of the scalar bilinear contribution.
From this we derive the exact expression of the generalized cusp anomaly for extreme ABJ ($\langle W \rangle \sim \exp\left( - \Gamma_{1/6}(\phi,\theta) \log\frac{\Lambda}{\epsilon} \right)$) \cite{Bianchi:2014laa}
\begin{equation}
\Gamma_{1/6}(\phi,\theta) = \lambda_1\, \frac{\sin\pi\lambda_2}{\pi} \left[ \cos \phi - \cos^2 \frac{\theta}{2} \right]\, \frac{\phi}{\sin \phi} + {\cal O}(\lambda_1^2) \,, \nonumber
\end{equation}
where $\phi$ is the deviation from the straight line configuration and $\theta$ is an internal angle in R-symmetry space.
Taking the coefficient of $\phi^2$ in the small angle expansion we find the exact Bremsstrahlung function \cite{Lewkowycz:2013laa}
\begin{equation}
B_{1/6} = \frac{\lambda_1\sin\pi\lambda_2}{2\pi} + {\cal O}(\lambda_1^2) \,.
\end{equation}

\section{Proof of the MOSS conjecture}

Finally we analyze the dilatation operator of ABJ in the $SU(2)\times SU(2)$ sector.
This was determined up to four loops in \cite{MOS1,MOS2,LMMSSST}, and its form in the extremal limit was conjectured in \cite{Minahan:2010nn} by Minahan, Ohlsson Sax and Sieg. We refer to this as the MOSS conjecture.
We prove this conjecture in two steps. First we determine the form of the dilatation operator to all loops up to a function of $\lambda_2$. Second we fix this function using the cusp anomalous dimension of the previous section.

We consider the dilatation operator ${\cal D}$ in superfield formalism, taking a chiral operator as the vacuum of a spin chain of asymptotic length $2L$.
The use of superfields perturbation theory allows to describe this sector in terms of organized structures which naturally emerge from the formalism itself. More specifically, one defines recursively the basis of chiral functions
\begin{align}
& \chi()=\{\},\quad \chi(a)=\{a\}-\chi(), \nonumber\\&
\chi(a,b)=\{a,b\}-\chi(a)-\chi(b)-\chi() \,,
\end{align}
and so on, as combinations of permutations $\{\}$ of fields, defined e.g. in \cite{Minahan:2010nn,Mauri:2013vd}.
These objects capture the nature of the chiral superpotential vertices, which are the only interactions that exchange flavour. 
Specifically, when one lists all the diagrams which contribute to the renormalization of operators, the chiral skeleton of a diagram (namely the chiral vertices and propagators on which vector interactions can be added) produces only one chiral function. The converse is not true as two different chiral skeletons may generate the same chiral function.

By construction, in any diagram the number of chiral and anti-chiral vertices is the same. Therefore, it is always possible to group them into adjacent pairs, connected by none, one, two or three chiral propagators. 
The last two possibilities are flavour-trivial, whereas the first two contribute to the chiral structures $\chi(1,2)$ and $\chi(1)$, forming effective eight and six-vertices. In this way we can unambiguously determine a one to one correspondence between chiral functions and effective chiral skeletons. 
Then, combining 
effective vertices, gives rise to chiral functions with higher degree, see \cite{Mauri:2013vd}. 

In terms of color contributions, however we combine multiple effective six and eight-vertices produces higher powers of $\lambda_1$. Moreover the effective eight-vertex alone already contributes with $\lambda_1^2\lambda_2^2$ \cite{LMMSSST}. Thus, in the extremal limit, the only leading chiral structure is the effective 6-vertex, appearing first at two loops.
Consequently, determining the full dilatation operator boils down to computing all the relevant flavour neutral, subleading in $\lambda_1$, corrections to the two-loop diagram. Hence at (even) loop order $2l$ the dilatation operator reads
$\mathcal{D}_{2l}\to G_{2l}\left(\chi(1)+\chi(2)\right)$, with $\chi(1)$ ($\chi(2)$) acting on odd (even) sites.
The function $G_{2l}=G_{2l}(\lambda_1,\lambda_2)$ is completely determined in terms of the $h$-function coefficients by imposing the magnon dispertion relation
\begin{equation}\label{eq:disprel}
E(p)=-\frac{1}{2}+\frac{1}{2}\sqrt{1+16 h^2(\lambda_1,\lambda_2)\sin^2\tfrac{p}{2}} \,.
\end{equation}
on the single magnon states. In the leading $\lambda_1$ limit only $G_{2l}\to-h_{2l}$ survives, since further products and powers of the $h$ coefficients $h^2(\lambda_1,\lambda_2)=\sum\limits_{l=1}^{\infty}h_{2l}$ are subleading. This means that we can resum the full dilatation operator in this limit to obtain
\begin{equation}
\mathcal{D}=L-h^2(\lambda_1,\lambda_2)\,\left(\chi(1)+\chi(2)\right) + {\cal O}(\lambda_1^2) \,.
\end{equation}
Hence, in the extremal limit the dilatation operator, at least within the $SU(2)\times SU(2)$ sector, is the same as that of two decoupled Heisenberg spin chains and hence the spectral problem is integrable to all loop orders. It is therefore natural to assume that the ABJ theory is integrable in this limit. Under this assumption we can use the integrability machinery of \cite{GV} and claim that the cusp anomalous dimension of ABJ is given by
\begin{equation}\label{eq:scalingABJ}
\Gamma_{cusp} = 4\, h^2(\lambda_1,\lambda_2) + {\cal O}(h^4) \,,
\end{equation}
where $h$ is the same interpolating function appearing in the magnon dispersion relation \eqref{eq:disprel}.
From both the Wilson loop and dilatation operator computations we see that $h^2$ contains a factor $\lambda_1\lambda_2$, therefore in the extremal limit only the first order in $h^2$ contributes in the perturbative expansion.
Thus, comparing \eqref{eq:scalingABJ} with \eqref{eq:cusp} we determine the $h$-function of extremal ABJ
\begin{equation}
h^2(\lambda_1,\lambda_2) = \lambda_1\, \frac{\sin\pi\lambda_2}{\pi} + {\cal O}(\lambda_1^2) \,,
\end{equation}
which concludes the proof of the MOSS conjecture.

\section*{Acknowledgements}

We thank Luca Griguolo, Andrea Mauri, Silvia Penati and Domenico Seminara for very useful discussions. The work of MB was supported in part by the Science and Technology Facilities Council Consolidated Grant ST/L000415/1 \emph{String theory, gauge theory \& duality}.

\bibliographystyle{apsrev4-1}

\bibliography{biblio}

\end{document}